\documentclass[twocolumn,english,aps,prl,superscriptaddress]{revtex4-1}
\usepackage{amsmath}
\usepackage{newtxmath}
\usepackage[latin9]{inputenc}
\usepackage{xcolor}
\usepackage{babel}
\usepackage{graphicx}
\usepackage[bookmarks=false,
 breaklinks=false,pdfborder={0 0 1},backref=false,colorlinks=true]
 {hyperref}
\hypersetup{
 citecolor=blue,linkcolor=blue}

\makeatletter
\allowdisplaybreaks[2]

\makeatother

\begin{document}
\title{Antiblockade Quantum Batteries}
\author{Guohui Dong}
\email{dongguohui@sicnu.edu.cn}

\affiliation{College of Physics and Electronic Engineering, Sichuan Normal University,
Chengdu 610068, China}
\author{Yao Yao}
\email{yaoyao_mtrc@caep.cn}

\affiliation{Microsystem and Terahertz Research Center, China Academy of Engineering
Physics, Chengdu 610200, China}
\begin{abstract}
Blockade, a phenomenon in which one particle hinders the transport
of another, constitutes a core research topic in mesoscopic and microscopic
physics. Benefiting from the unique single\nobreakdash-particle characteristics,
it underpins versatile applications spanning from quantum computation
to quantum metrology. Nevertheless, in the context of quantum energy\nobreakdash-storage
devices, i.e., quantum batteries (QBs), blockade triggered by the
intrinsic level splitting is in turn detrimental to energy accumulation.
Here we propose an antiblockade charging protocol enabled by deliberately
engineering a bright-dark interaction. Rooted in a manifold of equally-spaced
dressed-dark states, this elaborate configuration opens up a new charging
channel facilitating an antiblockade energy flow and outperforms the
Rabi-split blockade design with at least a 20-fold enhancement in
energy storage. Additionally, we further illustrate the achievement
of superextensive (quadratic) charging characteristic by operating
the system in the bad-cavity limit. In view of its inherent generality,
our scheme offers a universal and feasible pathway toward high-performance
quantum energy architectures.
\end{abstract}
\maketitle
\textit{Introduction}.---Nowadays, the enormous demands of renewable
energy resources, alongside the ongoing trend of device miniaturization,
have dramatically stimulated the development of energy storage techniques
in the microscopic domain, e.g., quantum batteries (QBs) \citep{Alicki2013,Hovhannisyan2013,Andolina2019,Zhang2019,Caravelli2020,Francica2020,Lu2021,Peng2021,Shi2022,Yang2023,Ahmadi2024,Campaioli2024,Lu2025,Yan2026,Ferraro2026,Barra2019,Santos2019,Quach2020,Santos2021,Carrasco2022,Downing2023,Yang2024,Song2024,Pokhrel2025,Cavaliere2025,Yang2026,Dias2026}.
Akin to the pronounced improvement of quantum protocols over their
classical counterparts in multiple tasks \citep{Gisin2007,Steane1998,Ladd2010,Preskill2018,Degen2017,Pirandola2018},
the quantum advantage of QBs has been investigated under diverse strategies
\citep{Campaioli2017,Binder2015,Ferraro2018,JuliaFarre2020,Hu2026}.
A key finding reveals that the speedup in a QB is limited by the maximum
energy jump in a single transition step \citep{Gyhm2022}, hiniting
at the fine-grained picture of energy transport. In this sense, scrutinizing
the behaviour of the QB from the perspective of microscopic energy
transfer, which remains largely underexplored, delivers a brand-new
viewpoint for unraveling its underlying mechanism and delineates clear
guidelines for the subsequent performance improvement.

Blockade constitutes a fundamental effect in particle-transport processes,
e.g., electron transport, photon propagation, and magnon transmission
\cite{scully1997quantum,Alhassid2000,Nazarov2009,Imamoglu1997,Wang2022}.
Arising from the nonlinear interactions or quantum interference within
the system, blockade permits only a single particle to pass at one
time. In other words, the presence of a first particle inhibits the
transport of a second. Benefiting from this unique single\nobreakdash-particle
characteristic, blockade enables versatile applications ranging from
quantum computation to quantum metrology. For instance, Coulomb blockade
lays the foundation for single\nobreakdash-electron transistors \cite{Nazarov2009,Kastner1992},
while photon blockade serves as a key principle for nonclassical single\nobreakdash-photon
sources \cite{Birnbaum2005,Couteau2023}.

However, within the framework of energy storage, the energy accumulation
of QBs would be naturally impeded by the blockade effect of the system
\citep{Tian1992,Brecha1999,Birnbaum2005}. As a representative example,
in a driven two-level QB, due to the unequal energy splittings across
discrete eigenstates \citep{Downing2023}, a drive resonant with the
transition from the ground state to single-excitation subspace becomes
detuned from subsequent higher-order transitions. This mismatch thus
blocks continuous energy accumulation in the QB. In parallel, beyond
the weak-driving regime, bistability, as well as hysteresis, originates
from the nonlinear dynamics of the system, where the steady state
of QBs relies on the sweep direction of driving frequency or intensity
\citep{Gripp1996,Gripp1997}. Therefore, fluctuations within the charging
scenario may cause a sudden jump of the QB from the upper (high-energy)
branch to the lower (low-energy) one and hence lead to diminished
energy storage. In aggregate, excitation blockade and dynamical bistability
severely degrade the overall performance of QBs with energy splitting.

In this work, we demonstrate an antiblockade QB leveraging an ensemble
of three-level cells. Unlike the traditional two-level framework,
a coherent coupling arises between a newly incorporated (dark) state
and the native excited (bright) state, inducing a series of equally-spaced
dressed-dark states with nonvanishing charger-mode components. Thus,
in contrast to the Rabi-split routes suffering from the blockade and
bistability, the dressed dark state unlocks a new charging channel
that manifests antiblockade and bistability-free charging merits.
Consequently, this emergent channel featuring inherent frequency robustness
outperforms the conventional blockade counterpart in energy storage
by at least one order of magnitude. As an additional insight, we also
illustrate the achievement of quadratic power scaling in our dissipative
QBs. Particularly, in the bad-cavity limit, the lossy charger mode
effectively provides a collective decay channel for the QB with the
relaxation rate proportional to the QB size. Hence, together with
the linear energy storage, the charging power exhibits superextensive
behaviour, analogous to the collective emission burst observed in
superradiance \citep{Dicke1954,opensystem2002,Meiser2009,Dong2023,Dong2025a}.
Feasible on current experimental setups such as\textbf{ }diamond nitrogen-vacancy
(NV) centers and circuit quantum electrodynamics (QED), our proposal
offers fresh physical insights into the energy transfer and storage
mechanisms in QBs and may subsequently facilitate the development
of quantum energy techniques.

\begin{figure}[t]
\begin{centering}
\includegraphics[scale=0.8]{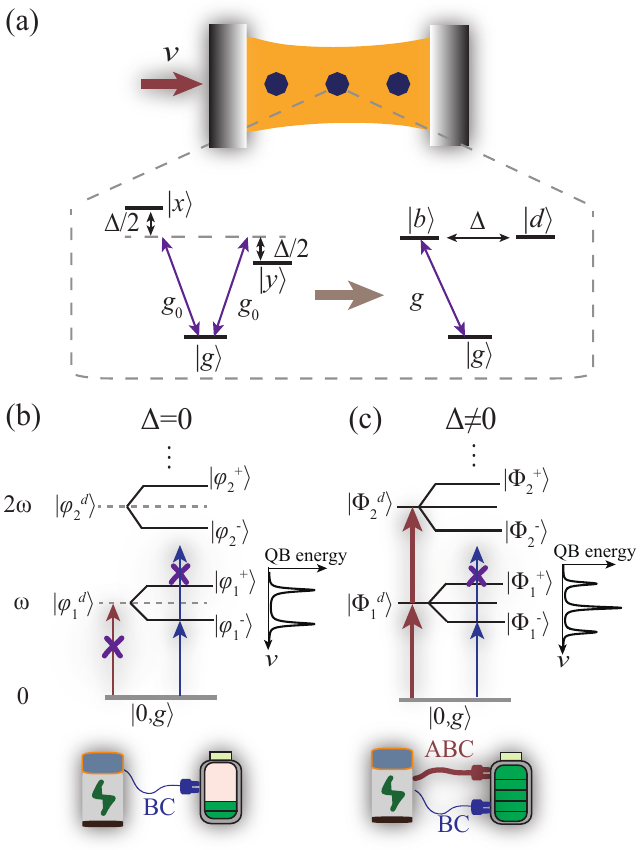}
\par\end{centering}
\caption{(a) The schematic diagram of the collectively charged V-type three-level
QB. In the bright and dark basis, only the bright state $\left|b\right\rangle $
couples with the charger mode. The energy gap $\Delta$ induces an
effective interaction between the bright $\left|b\right\rangle $
and dark $\left|d\right\rangle $ states. (b) The energy diagram of
the single-cell degenerate case ($\Delta=0$). The QB can only be
charged via the Rabi-split blockade channel. (c) The energy diagram
of the single-cell non-degenerate case ($\Delta\protect\neq0$). The
bright-dark coupling $\Delta$ opens up a new antiblockade channel.}\label{fig:1}
\end{figure}

\textit{Dark-state }QB.---Here we explore a collection of V-type
three-level cells collectively charged by a bosonic mode via coherent
driving (Fig. \ref{fig:1}(a)). In the charging process, the Hamiltonian
of the entire system reads ($\hbar=1$) 
\begin{align}
\hat{H} & =\omega_{c}\hat{c}^{\dagger}\hat{c}+\sum^{N}_{i=1}\left[\left(\omega_{0}+\frac{\Delta}{2}\right)\hat{\sigma}^{i}_{xx}+\left(\omega_{0}-\frac{\Delta}{2}\right)\hat{\sigma}^{i}_{yy}\right]\nonumber \\
 & +\frac{g_{0}}{2}\sum^{N}_{i=1}\left[\left(\hat{\sigma}^{i}_{xg}+\hat{\sigma}^{i}_{yg}\right)\hat{c}+\mathrm{H.c.}\right]+\left(\xi\hat{c}^{\dagger}e^{-i\nu t}+\mathrm{H.c.}\right),
\end{align}
where $\hat{c}$ represents the annihilation operator of the charger
mode with frequency $\omega_{c}$. $\hat{\sigma}^{i}_{mn}\equiv\vert m\rangle^{i}\langle n\vert$
($m,n=x,y,g$ and $i\in\{1,2,\cdots,N\}$) corresponds to the operator
of the $i$th QB cell with excited states $\left|x\right\rangle $
and $\left|y\right\rangle $ and ground state $\left|g\right\rangle $.
$\Delta$ characterizes the energy gap between two excited states
($\omega_{0}\pm\Delta/2$) and can be manipulated by external signal.
For instance, the degeneracy present in levels of atoms or NV centers
can be lifted via Zeeman effect \citep{scully1997quantum,Winchester2017,sakurai2021modern,Yin2015}.
The charger couples with both two transitions in the QB with homogeneous
Rabi frequency $g_{0}$ (assumed to be real). A coherent driving field
with frequency $\nu$ and relative strength $\xi$ delivers energy
to the QB through the charger \citep{Farina2019,Downing2023}. For
simplicity, we restrict our analysis to a resonant scenario ($\omega_{c}=\omega_{0}\equiv\omega$)
throughout this work.

By defining the bright and dark states $\left|b\right\rangle \equiv\left(\left|x\right\rangle +\left|y\right\rangle \right)/\sqrt{2}$
and $\left|d\right\rangle \equiv\left(\left|x\right\rangle -\left|y\right\rangle \right)/\sqrt{2}$,
the Hamiltonian in the interaction picture is rewritten as 
\begin{align}
\hat{H}_{I} & =\delta\hat{c}^{\dagger}\hat{c}+\sum^{N}_{i=1}\left[\delta\left(\hat{\sigma}^{i}_{bb}+\hat{\sigma}^{i}_{dd}\right)+\frac{\Delta}{2}\left(\hat{\sigma}^{i}_{bd}+\hat{\sigma}^{i}_{db}\right)\right]\nonumber \\
 & +\frac{g}{2}\sum^{N}_{i=1}\left(\hat{\sigma}^{i}_{bg}\hat{c}+\mathrm{H.c.}\right)+\left(\xi\hat{c}^{\dagger}+\mathrm{H.c.}\right),\label{eq:bd hamiltonian}
\end{align}
where $\delta\equiv\omega-\nu$ ($g\equiv\sqrt{2}g_{0}$) stands for
the QB-drive detuning (QB-charger coupling strength). In spite of
the decoupling of the dark state from the charger, the energy gap
$\Delta$ induces an effective interaction between the bright and
dark states in the new basis.

This bright-dark coupling is the salient characteristic of our proposal.
Roughly speaking, in the absence of $\Delta$, the three-level QB
recovers to a two-level situation where the dark state is genuinely
``dark'' (isolated from other modes). In this case, the energy flow
in the QB would be inevitably blocked due to the detuning of the unevenly-spaced
Rabi-split transitions against the driving frequency (Fig. \ref{fig:1}(b)).
By contrast, a nonvanishing $\Delta$ engenders a manifold of equally-spaced
dark eigenstates (i.e., $E^{d}_{n}=n\delta$) dressed with a charger
component \citep{Winchester2017,Dong2020,Zhang2021}. In essence,
this bright-dark interaction opens up a novel antiblockade channel
(ABC) beyond the blockade channel (BC), see Fig. \ref{fig:1}(c) and
Sec. S1 in \citep{SM}.

As two illustrative figures of merit in QBs, we explore the performance
of our scheme with the energy storage and mean charging power 
\begin{align}
\mathcal{E}\left(t\right) & =\omega\sum^{N}_{i=1}\mathrm{Tr}\left[\hat{\rho}\left(t\right)\left(\hat{\sigma}^{i}_{bb}+\hat{\sigma}^{i}_{dd}\right)\right],\\
\mathcal{P}\left(t\right) & =\frac{\mathcal{E}\left(t\right)}{t},
\end{align}
where $\hat{\rho}(t)$ is the reduced density matrix of the QB.

\begin{figure}[t]
\begin{centering}
\includegraphics[scale=0.2]{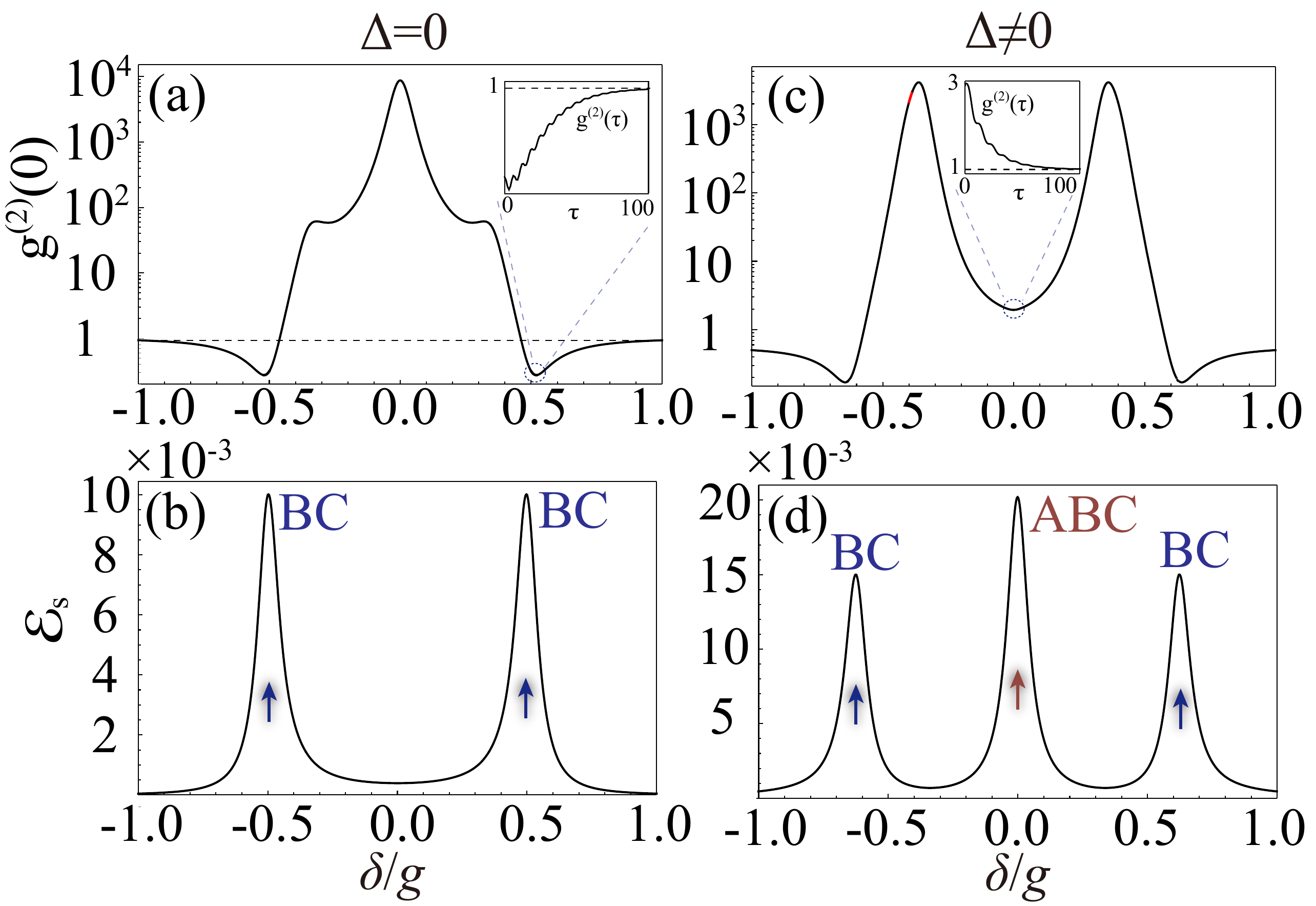}
\par\end{centering}
\caption{The spectra of the equal-time second-order correlation function $g^{(2)}(0)$
and steady-state energy storage $\mathcal{E}_{\mathrm{s}}$ for $\Delta=0$
(a,b) and $\Delta=0.7g$ (c,d). The insets in (a) and (c) show the
second-order correlation function $g^{(2)}(\tau)$ versus the time
delay $\tau$. Here $\mathcal{C}_{0}\equiv g^{2}/\kappa\gamma=100$,
$\kappa/\gamma=1$, and $\xi=0.01g$.}\label{fig:2}
\end{figure}

\textit{Antiblockade charging channel}.---Without loss of generality,
here we analytically exhibit the charging mechanism of our QB for
the case of $N=1$. Despite its simplicity, the single-cell design
captures the dominant physics while circumventing cumbersome notation
and lengthy derivations \citep{Andolina2018,Joshi2022,Hu2022,Zhu2023,Song2025}.
In fact, a more practical scheme should go beyond the Hamiltonian
(unitary) paradigm to incorporate the unavoidable noise effect \citep{Barra2019,Farina2019}.
Therefore, the total system can be described by an effective non-Hermitian
Hamiltonian $\hat{H}_{\mathrm{eff}}=\hat{H}_{\mathrm{eff,}0}+\hat{H}_{\mathrm{eff,}1}$
\begin{align}
\hat{H}_{\mathrm{eff,}0} & =\left(\delta-i\frac{\kappa}{2}\right)\hat{c}^{\dagger}\hat{c}+\left(\delta-i\frac{\gamma}{2}\right)\left(\hat{\sigma}_{bb}+\hat{\sigma}_{dd}\right)\nonumber \\
 & +\frac{\Delta}{2}\left(\hat{\sigma}_{bd}+\hat{\sigma}_{db}\right)+\frac{g}{2}\left(\hat{\sigma}_{bg}\hat{c}+\mathrm{H.c.}\right),\label{eq:H0}\\
\hat{H}_{\mathrm{eff,}1} & =\xi\hat{c}^{\dagger}+\mathrm{H.c.},\label{eq:H1}
\end{align}
where $\kappa$ ($\gamma$) represents the decay rate of the charger
(QB cell).

Obviously, a small driving term ($\hat{H}_{\mathrm{eff,}1}$) would
cause a slight correction to the eigensolutions of the excitation-conserved
Hamiltonian $\hat{H}_{\mathrm{eff,}0}$. Meanwhile, owing to the noise-induced
negative imaginary part in the eigenenergies, a QB initially in excited
states will gradually dissipates, leading to a unique steady state,
i.e., the corrected ground state (see Sec. S1 in \citep{SM}) 
\begin{align}
\left|G\right\rangle  & \simeq\left|0,g\right\rangle +\sum_{j=\pm,d}\left(\alpha_{j}\left|\Phi^{j}_{1}\right\rangle +\beta_{j}\left|\Phi^{j}_{2}\right\rangle \right),\label{eq:groundstate}
\end{align}
where $\left|n,p\right\rangle \equiv\left|n\right\rangle _{C}\otimes\left|p\right\rangle _{B}$
$\left(p=g,b,d\right)$ denotes the charger-QB product state. $\left|\Phi^{j}_{n}\right\rangle $
is the eigenstate of $\hat{H}_{\mathrm{eff,}0}$ with $n$ excitations.
$\pm$ ($d$) indicates the Rabi-split (dark) eigenstate. It is worth
emphasizing that the bright-dark coupling elicits a nonzero weight
of dark eigenstate $\left|\Phi^{d}_{n}\right\rangle $ which is dressed
with the charger mode in Eq. (\ref{eq:groundstate}). Following the
convention in quantum optics, below we label the Rabi-split and dark
eigenstates the dressed bright and dressed dark states, respectively.

In the steady state, the energy stored in the QB can be recast as
\begin{align}
\mathcal{E}_{\mathrm{s}} & \equiv\underset{t\rightarrow\infty}{\lim}\left(\mathcal{E}\left(t\right)\right)\nonumber \\
 & \simeq\frac{\omega}{2}\left|\alpha_{+}+\alpha_{-}\right|^{2}+\frac{\omega}{2}\left|\frac{\Delta\left(\alpha_{+}-\alpha_{-}\right)}{g}-\sqrt{2}\alpha_{d}\right|^{2},\label{eq:EB}
\end{align}
where only the lowest-order excitations are retained. As shown in
Eq. (\ref{eq:EB}), both the dressed bright and dark states contribute
to the energy storage in the QB. To explicitly uncover the underlying
physics in the charging mechanism, we further scrutinize the statistical
behaviour of the steady-state excitations. Especially, the equal-time
second-order correlation function in charger can be analytically expressed
as \citep{opensystem2002,scully1997quantum} 
\begin{align}
g^{(2)}(0) & =\frac{\left\langle G\right|\hat{c}^{\dagger}\hat{c}^{\dagger}\hat{c}\hat{c}\left|G\right\rangle }{\left\langle G\right|\hat{c}^{\dagger}\hat{c}\left|G\right\rangle ^{2}}\nonumber \\
 & \simeq\frac{4\left|\beta_{+}-\beta_{-}+\beta_{d}\frac{\Delta}{g}\right|^{2}}{\left|\alpha_{+}-\alpha_{-}+\alpha_{d}\frac{\sqrt{2}\Delta}{g}\right|^{4}}.\label{eq:g20}
\end{align}

In the absence of the bright-dark coupling ($\Delta=0$), the charging
of the QB is drastically suppressed by excitation blockade. Specifically,
energy can only be injected into the QB via the dressed bright state
($\alpha_{d}=\beta_{d}=0$), yielding distinct Rabi peaks in the excitation
spectrum \citep{Yang2026}. Nevertheless, owing to the uneven energy
spacing of the dressed bright levels \citep{Downing2023}, a driving
field resonant with the transition from the zero-excitation state
to the single-excitation subspace would inevitably be detuned from
higher transitions, which severely hinders successive energy accumulation
(Fig. \ref{fig:1}(b)). Further investigation on the statistical properties
of excitations in the charger confirms the theoretical analysis. As
revealed in Eq. (\ref{eq:g20}), the second-order correlation function
of the charger for $\Delta=0$ satisfying $g^{(2)}\left(0\right)\simeq0$
and $g^{(2)}\left(0\right)<g^{(2)}\left(\tau\right)$ signifies a
sub-Poissonian distribution and excitation blockade, i.e., antibunching
\citep{scully1997quantum,opensystem2002} in the BC (Figs. \ref{fig:2}(a)
and \ref{fig:2}(b)).

In sharp contrast, it turns out that the blockade can be lifted when
the dark state is coupled with bright state ($\Delta\neq0$). Particularly,
a new charging peak stemming from the dressed dark states ($\alpha_{d},\beta_{d}\neq0$)
emerges once the drive resonates with the transition $\left|\Phi^{d}_{1}\right\rangle -\left|\Phi_{0}\right\rangle $
(Fig. \ref{fig:1}(c)). Crucially, the equal spacing of the dressed
dark states removes the obstacle of off-resonant driving in subsequent
excitations. The second-order correlation function $g^{(2)}\left(0\right)\simeq g^{2}\left|\beta_{d}/\alpha^{2}_{d}\right|^{2}/\Delta^{2}>g^{(2)}\left(\tau\right)\geq1$
demonstrates super-Poissonian statistics and antiblockade (i.e., bunching)
for excitations propagating via the ABC (Figs. \ref{fig:2}(c) and
\ref{fig:2}(d)). It should be stressed that the above analytical
results are in excellent agreement with numerical master-equation
simulations. Moreover, beyond the single-cell setup, the antiblockade
feature rooted in the dressed dark states persists in the collective
charging configurations (see Secs. S2 and S3 in \citep{SM}).

\begin{figure}[t]
\begin{centering}
\includegraphics[scale=0.28]{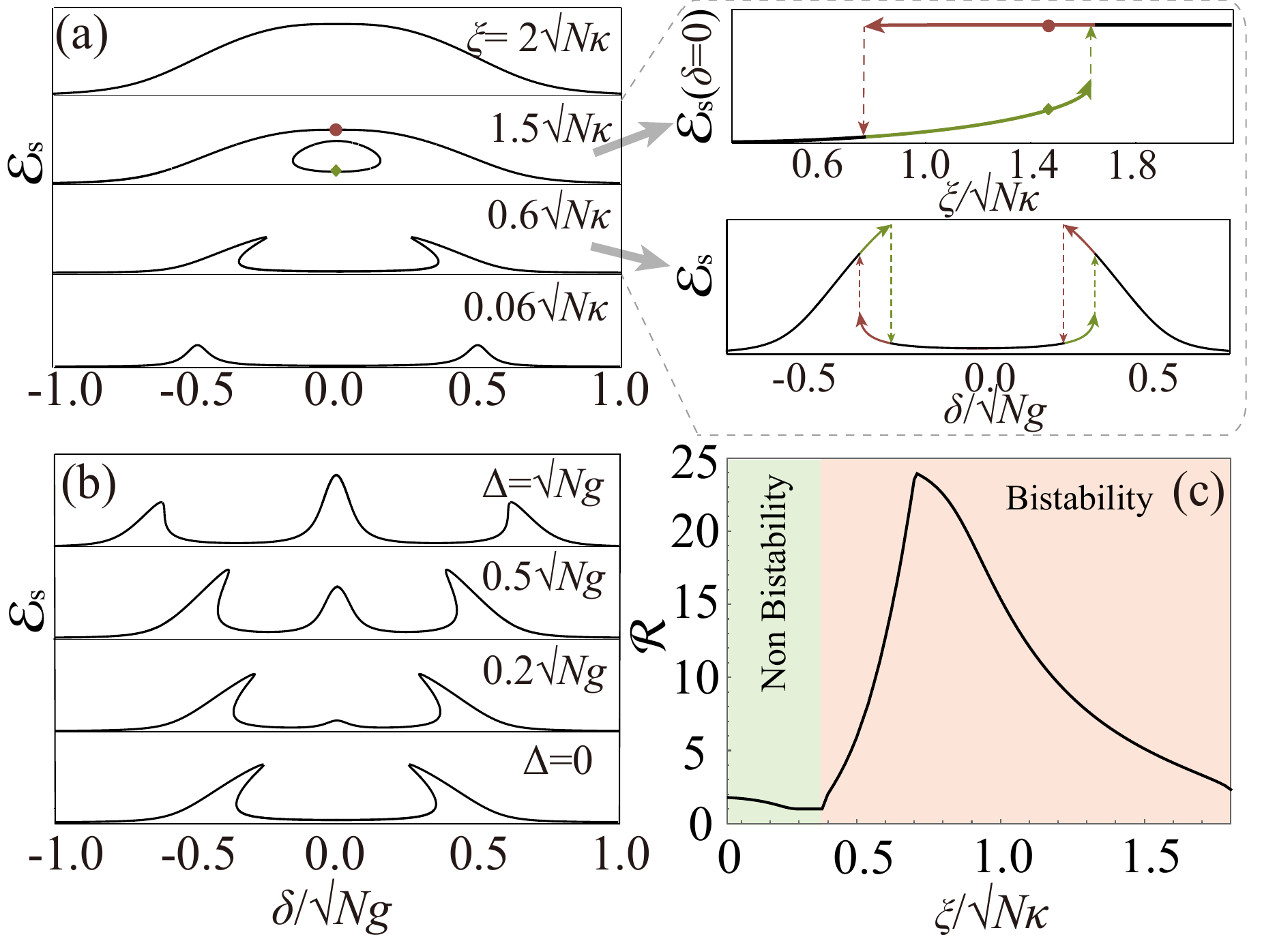}
\par\end{centering}
\caption{(a) Left panel: The collective charging spectra for a series of driving
strength with $\Delta=0$. Right panel: The driving-intensity and
-frequency hysteresis. The red circle and green diamond in the left
panel for $\xi=1.5\sqrt{N}\kappa$ correspond to those in the intensity
hysteresis (top-right panel). (b) The collective charging spectra
for multiple $\Delta$. (c) The capacity advantage of the ABC over
the BC. Here $\Delta=0$ for (a), $\xi=0.6\sqrt{N}\kappa$ for (b),
$\Delta=\sqrt{N}g$ for (c), $N=10^{6}$, $\mathcal{C}_{0}\equiv g^{2}/\kappa\gamma=10^{-4}$,
and $\kappa/\gamma=1$.}\label{fig:3}
\end{figure}

\textit{ABC superiority.}---Notably, the ABC resonance peak is intrinsically
endowed with frequency robustness. Originated from the nonlinear dynamics
of the system, two Rabi sidebands initially deform toward the center
with the increase of the driving strength and eventually merge into
a single peak (Fig. \ref{fig:3}(a)), which triggers the appearance
of bistability in the BC (see Sec. S4 in \citep{SM}).\textbf{ }On
the contrary, the resonance peak of the ABC maintains at the center
of the spectrum ($\delta=0$), irrespective of variations in the driving
intensity or bright-dark coupling (Fig. \ref{fig:3}(b)). This frequency
stability of the ABC peak significantly relaxes the experimental requirements
of the driving field and thus highlights the feasibility of our scenario.

More importantly, the ABC gives rise to prominent energy-storage capability
over the BC. As elucidated in the right panel of Fig. \ref{fig:3}(a),
the bistability of the BC leads to a frequency or intensity hysteresis
depending on the driving strength \citep{Gripp1996,Gripp1997}. Therefore,
fluctuations during the driving process would cause a sudden jump
of the QB from the upper branch to the lower one of the bistability
curve, which heavily hinders the energy storage of the QB. Conversely,
the peak of the ABC facilitates a gradual increase with the driving
strength (Fig. \ref{fig:3}(b)), harvesting a significant promotion
of energy storage. To quantitatively characterize the outperformance
of the ABC, we define a capacity advantage $\mathcal{R}=\mathcal{E}_{A}/\mathcal{E}_{B}$
where $\mathcal{E}_{A\left(B\right)}$ stands for the ABC (BC) peak
energy. Note that within the bistable region, $\mathcal{E}_{B}$ corresponds
to the energy of the lower branch of the BC. As illustrated in Fig.
\ref{fig:3}(c), the ABC enables a 24-fold boost in energy storage
relative to the BC across the bistability window. It is worth pointing
out that this pronounced improvement just arises in the hysteresis
regime and hence marks the upper threshold of the achievable superiority.

\textit{Quadratic scaling of charging power}.---In collective frameworks,
one of the most crucial figures of merit is the charging scaling,
especially the power. It has been demonstrated that the charging power
markedly benefits from the global nature of charging operations and
manifests at most quadratic behaviour \citep{Gyhm2022}. Notably,
in the noise-free collective charging proposal, the work in \citep{Ferraro2018}
achieve $\sim N{}^{3/2}$ charging power which inherits from the $\sqrt{N}$
enhancement of the effective QB-charger interaction. For organic molecules
embedded in a microcavity, light can be captured and converted into
an electric current superextensively \citep{Hymas2026}. Here, we
examine the charging dynamics in our ABC scenario ($\delta=0$) and
illuminate the quadratic power scaling, approaching the theoretical
upper limit.

In our collective protocol, the mean-field equations of the system
can be linearized as 
\begin{align}
\frac{\partial\left\langle \hat{c}\right\rangle }{\partial t} & =-\frac{\kappa}{2}\left\langle \hat{c}\right\rangle -i\frac{\sqrt{N}g}{2}\left\langle \hat{B}\right\rangle -i\xi,\label{eq:linear equation a}\\
\frac{\partial\left\langle \hat{B}\right\rangle }{\partial t} & =-\frac{\gamma}{2}\left\langle \hat{B}\right\rangle -i\frac{\sqrt{N}g}{2}\left\langle \hat{c}\right\rangle -i\frac{\Delta}{2}\left\langle \hat{D}\right\rangle ,\label{eq:linear equation B}\\
\frac{\partial\left\langle \hat{D}\right\rangle }{\partial t} & =-\frac{\gamma}{2}\left\langle \hat{D}\right\rangle -i\frac{\Delta}{2}\left\langle \hat{B}\right\rangle ,\label{eq:linear equation D}
\end{align}
where $\left\langle \hat{O}\right\rangle $ denotes the expectation
of the operator $\hat{O}$. $\hat{B}(\hat{D})$ is the bosonic annihilation
operator describing the collective bright (dark) excitations (see
Sec. S5 in \citep{SM}). Since the driving strength $\xi\equiv\sqrt{\kappa P_{\mathrm{in}}/2\omega}$
scales with the square root of the external field power $P_{\mathrm{in}}$
\citep{Tieri2015,Dong2020}, for a fair comparison of the charging
behaviour across distinct QB sizes, one should constrain the driving
strength with an equal mean energy cost $\left|\xi\right|^{2}/N$.
Hence, the steady-state QB energy ($\mathcal{E}_{\mathrm{s}}=\omega\left|\left\langle \hat{B}\right\rangle _{\mathrm{s}}\right|^{2}+\omega\left|\left\langle \hat{D}\right\rangle _{\mathrm{s}}\right|^{2}\propto\left|\xi\right|^{2}$)
reveals a linear battery capacity. Accordingly, a charging time inversely
proportional to the QB size will yield a quadratic power scaling.

In the bad-cavity regime where the charger dissipates much faster
than the QB ($\kappa\gg\gamma$), the cavity mode can instantaneously
track the dynamics of other slowly evolving variables and thus be
adiabatically eliminated \citep{Brion2007,Liu2020}. Namely, the cavity
mode in Eqs. (\ref{eq:linear equation a})-(\ref{eq:linear equation D})
effectively supplies a collective decay channel for the QB \citep{Meiser2010}
\begin{align}
\frac{\partial\left\langle \hat{B}\right\rangle }{\partial t} & =-\frac{1}{2}\left(\gamma+\frac{Ng^{2}}{\kappa}\right)\left\langle \hat{B}\right\rangle -i\frac{\Delta}{2}\left\langle \hat{D}\right\rangle -\frac{\sqrt{N}g}{\kappa}\xi,\label{eq:linear equation B-1}\\
\frac{\partial\left\langle \hat{D}\right\rangle }{\partial t} & =-\frac{\gamma}{2}\left\langle \hat{D}\right\rangle -i\frac{\Delta}{2}\left\langle \hat{B}\right\rangle .\label{eq:linear equation D-1}
\end{align}

As exhibited in Eq. (\ref{eq:linear equation B-1}), the effective
dissipation rate of the bright excitation scales linearly with the
QB size. Therefore, combined with the extensive energy capacity, the
collective charging QB in the bad-cavity limit delivers a power proportional
to $N^{2}$, resembling the collective emission burst in optical superradiance
\citep{Dong2023,Dong2025a}. For representative parameters, e.g.,
$\xi=0.1\sqrt{N}\kappa$, $\Delta=\sqrt{N}g$, the average stored
energy and charging power of the QB read $\mathcal{E}_{\mathrm{s}}\simeq0.04N\omega$
and $\mathcal{P}_{\mathrm{s}}\sim0.04N^{2}\omega\gamma\mathcal{C}_{0}$
where $\mathcal{C}_{0}\equiv g^{2}/\kappa\gamma$ is the single-cell
cooperativity parameter. Note that the self-consistency of the adiabatic
elimination ($N\mathcal{C}_{0}\gg1$, $\kappa^{2}>Ng^{2}$ and $N<\Delta\kappa/g^{2}$)
effectively imposes a confining constraint on the QB size, i.e., $N_{\mathrm{min}}\equiv\kappa\gamma/g^{2}\ll N<\mathrm{min}\left(\kappa^{2}/g^{2},\Delta\kappa/g^{2}\right)\equiv N_{\mathrm{max}}$.
Below $N_{\mathrm{min}}$, the collective enhanced decay rate $Ng^{2}/\kappa$
in Eq. (\ref{eq:linear equation B-1}) fails to dominate the charging
process, while above $N_{\mathrm{max}}$ the dynamical behaviour of
the QB is mainly determined by the intrinsic relaxation of the charger
with the charging time $\sim1/\kappa$. Consequently, beyond the superradiant
regime, the quadratic charging power reverts to an extensive power
scaling.

\begin{figure}[t]
\begin{centering}
\includegraphics[scale=0.2]{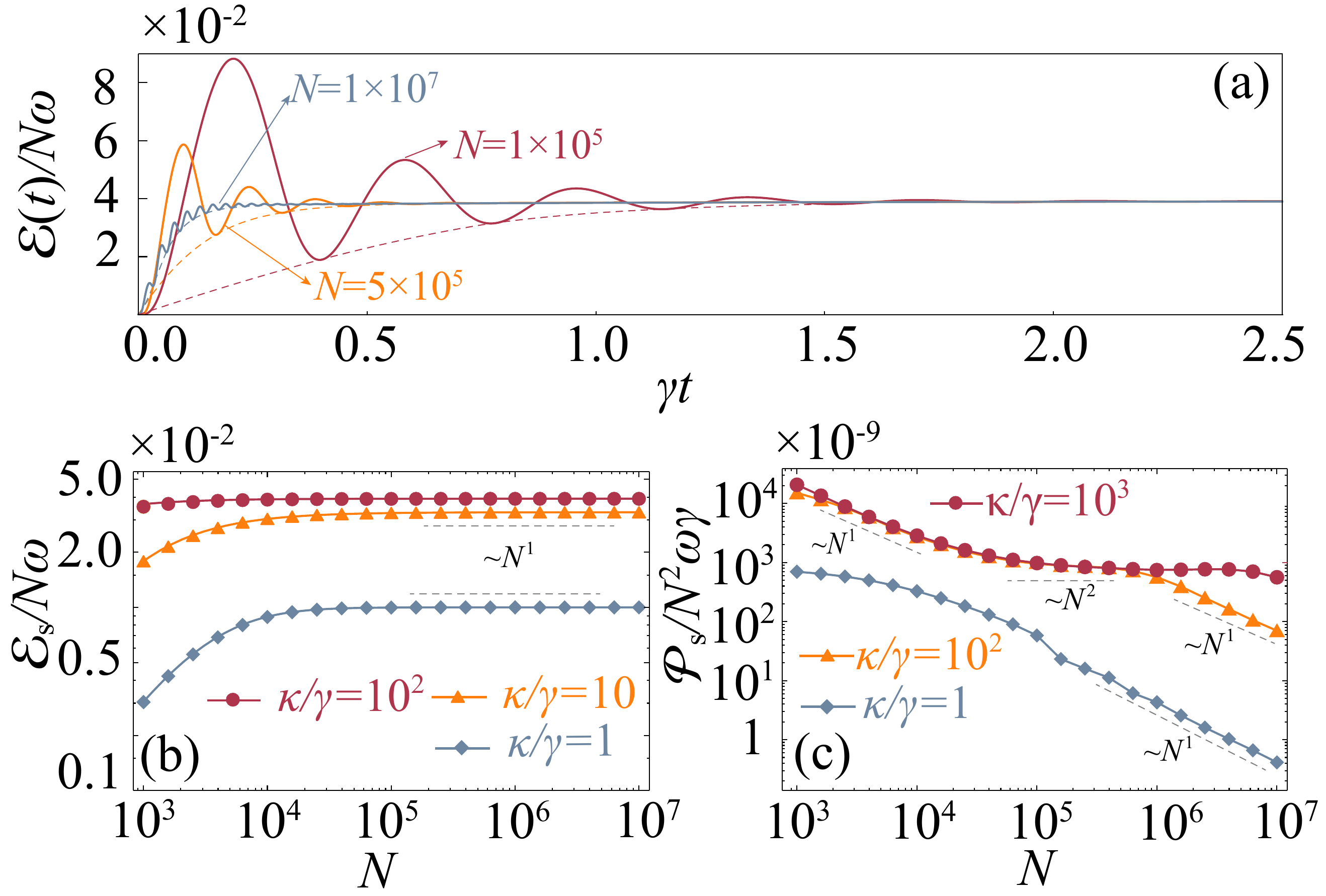}
\par\end{centering}
\caption{(a) The energy in the QB versus time for representative QB size,
e.g., $N=10^{5},5\times10^{5}$, and $10^{7}$ (magenta, orange, and
blue). (b) The steady-state energy in the QB versus $N$ for $\kappa/\gamma=10^{2},10,$
and 1 (magenta line with circles, orange line with triangles, and
blue line with diamonds). (c) The steady-state charging power in the
QB versus $N$ for $\kappa/\gamma=10^{3},10^{2},$ and 1 (magenta
line with circles, orange line with triangles, and blue line with
diamonds). Here $\kappa/\gamma=10^{2}$ for (a), $\xi=0.1\sqrt{N}\kappa$
for (b,c), $\Delta=\sqrt{N}g$, and $\mathcal{C}_{0}\equiv g^{2}/\kappa\gamma=10^{-4}$.}\label{fig:4}
\end{figure}

The superradiant charging performance under the bad-cavity limit is
well displayed in Fig. \ref{fig:4}. As elucidated in Figs. \ref{fig:4}(a)
and \ref{fig:4}(b), larger QBs feature faster dynamics (see the lower
envelope of the curve) while sustaining identical (size-independent)
steady-state energy per cell, which implies potential superextensive
power. Specially, despite the linear scaling for the case $\kappa/\gamma=1$
(no ``bad cavity''), a near quadratic power scaling emerges when
one pushes the system down toward the bad-cavity regime (Fig. \ref{fig:4}(c)).
As analyzed above, a deeper penetration into the bad-cavity limit
would give rise to a broader size region with quadratic scaling (see
Sec. S5 in \citep{SM}). It is noteworthy that since the collective
enhanced decay channel stems from the rapid relaxation of the charger,
the superlinear power is attained at the cost of substantial dissipation
and degraded efficiency \citep{Pokhrel2025}.

\textit{Discussions and conclusions}.---Current experimental advances
in solid-state quantum technology, cavity and circuit QED have offered
a firm foundation for the feasibility of our proposal. In particular,
our dark-state scheme can be realized with hybrid quantum platforms,
including NV centers coupled with a transmission line resonator (TLR)
as well as realistic atoms or organic molecules embedded in an optical
cavity, where the bright-dark coupling can be tailored via external
fields \citep{Yang2011,Winchester2017,Tao2015,Skulte2021,Fan2023,Hymas2026,Yang2022,Norcia2016}.
As a paradigmatic example, we showcase the implementation of our protocol
using NV-center ensembles within circuit QED configurations. Benefiting
from the $\sqrt{N}$ enhancement of the collective coupling, the strong-coupling
regime, where the energy exchange between the QB and charger prevails
over the dissipation dynamics, would be conveniently accessed \citep{Yang2011,Tao2015}.
Concretely, the coupling strength between the NV centers and TLR can
reach several tens of MHz \citep{Kubo2010}. A static magnetic field
around 10G suffices to generate a typical bright-dark coupling strength
($\Delta$ in the tens of MHz range). The long coherence time of NV
centers and high quality factor of the TLR, i.e., $\gamma$ and $\kappa$
on the order of hundreds of kHz \citep{Yang2011,Tao2015,Song2025},
uphold the strong-coupling requirements of our charging strategy ($N\mathcal{C}_{0}\gg1$).
Moreover, the quadratic charging scaling (the bad-cavity limit) can
be acquired by mitigating the relaxation of the spin in diamond through
cooling and isotopically engineering techniques \citep{Cambria2023,Balasubramanian2009}.

In summary, we have demonstrated an antiblockade QB scenario utilizing
a collection of three-level QB cells. Distinct from the unevenly-spaced
Rabi-split eigensturcture, the bright-dark coupling hosts a manifold
of equally-spaced dressed-dark states, indicative of the emergence
of a new charging dynamics. Specifically, this dressed-dark channel
enables an antiblockade energy flow with inherent frequency robustness.
Moreover, by circumventing the bistability regime, our dark-state
scheme outperforms the conventional Rabi-split design, yielding at
least a 20-fold promotion in energy storage compared to the BC. By
operating the system in the bad-cavity regime, we further illustrate
the achievement of superextensive charging power. Notably, rooted
in the harmonic feature of the eigenenergy ladder, the antiblockade
charging strategy we proposed here is achievable in diverse protocols
beyond the three-level bright-dark interaction design in our work,
for instance a quadratic driving scheme (see Sec. S6 in \citep{SM}).
Feasible on current experimental platforms such as NV centers and
circuit QED, our proposal offers fresh physical perspectives into
the energy transfer and storage mechanisms of open QBs and may remarkably
facilitate the development of realistic quantum energy techniques
in the future.

\textit{Acknowledgments}.---We thank Prof. Peng Zhang for fruitful
discussions. G.D. is supported by the Sichuan Science and Technology
Program (Grant No. 2026NSFSC0730) and National Natural Science Foundation
of China (Grant No. 12205211). Y.Y. is supported by the National Natural
Science Foundation of China (Grant No. 12175204).

\textit{Data availability}.---The data are available upon reasonable
request from the authors.

\end{document}